\documentclass[twocolumn,prl,superscriptaddress]{revtex4}
\usepackage{amsmath,amssymb,amsfonts,mathrsfs}
\usepackage{bm}
\usepackage{graphicx}
\usepackage{hyperref}
\usepackage{amsbsy}
\usepackage{dcolumn}
\usepackage{lipsum}

\hypersetup{
	colorlinks = true,
	citecolor = red, 
	linkcolor = blue 
}

\begin{document}
\title{Experimental Implementation of Noncyclic and Nonadiabatic Geometric Quantum Gates in a Superconducting Circuit}

\author{Zhuang Ma}
\thanks{These three authors contributed equally to this work.}
\affiliation{National Laboratory of Solid State Microstructures, School of Physics, Nanjing University, Nanjing 210093, China}
\author{Jianwen Xu}
\thanks{These three authors contributed equally to this work.}
\affiliation{National Laboratory of Solid State Microstructures, School of Physics, Nanjing University, Nanjing 210093, China}
\author{Tao Chen}
\thanks{These three authors contributed equally to this work.}
\affiliation{Guangdong Provincial Key Laboratory of Quantum Engineering and Quantum Materials, and School of Physics and Telecommunication Engineering, South China Normal University, Guangzhou 510006, China}
\author{Yu Zhang}
\affiliation{National Laboratory of Solid State Microstructures, School of Physics, Nanjing University, Nanjing 210093, China}
\author{Wen Zheng}
\affiliation{National Laboratory of Solid State Microstructures, School of Physics, Nanjing University, Nanjing 210093, China}
\author{Dong Lan}
\affiliation{National Laboratory of Solid State Microstructures, School of Physics, Nanjing University, Nanjing 210093, China}
\affiliation{Hefei National Laboratory, Hefei 230088, China}
\author{Zheng-Yuan Xue} 
\email[]{zyxue83@163.com}
\affiliation{Guangdong Provincial Key Laboratory of Quantum Engineering and Quantum Materials, and School of Physics and Telecommunication Engineering, South China Normal University, Guangzhou 510006, China}
\author{Xinsheng Tan}
\email[]{tanxs@nju.edu.cn}
\affiliation{National Laboratory of Solid State Microstructures, School of Physics, Nanjing University, Nanjing 210093, China}
\affiliation{Hefei National Laboratory, Hefei 230088, China}
\author{Yang Yu}
\affiliation{National Laboratory of Solid State Microstructures, School of Physics, Nanjing University, Nanjing 210093, China}
\affiliation{Hefei National Laboratory, Hefei 230088, China}
\date{\today}

\begin{abstract}
 Quantum gates based on geometric phases possess intrinsic noise-resilience features and therefore attract much attention. However, the implementations of previous geometric quantum computation typically require a long pulse time of gates. As a result, their experimental control inevitably suffers from the cumulative disturbances of systematic errors due to excessive time consumption. Here, we experimentally implement a set of noncyclic and nonadiabatic geometric quantum gates in a superconducting circuit, which greatly shortens the gate time. And also, we experimentally verify that our universal single-qubit geometric gates are more robust to both the Rabi frequency error and qubit frequency shift-induced error, compared to the conventional dynamical gates, by using the randomized benchmarking method.
 Moreover, this scheme can be utilized to construct two-qubit geometric operations, while the generation of the maximally entangled Bell states is demonstrated. Therefore, our results provide a promising routine to achieve fast, high-fidelity, and error-resilient quantum gates in superconducting quantum circuits.
\end{abstract}

\maketitle

The superconducting quantum circuit is one of the promising candidates for future large-scale quantum computation \cite{Kjaergaard2020} due to its high controllability and scalability. At this stage, the major obstacle is relatively short coherence time and experimental perturbations, which demand speeding up quantum operations and improving the robustness against errors under the experimental controls in superconducting quantum circuits. Therefore, with their intrinsic noise-resilience features, the gates induced by geometric phases \cite{Victor1984, Zanardi1999, Jones2000}, attainable in superconducting systems, are highly anticipated.

The geometric phases depend only on the global properties of their evolution paths, so that they can be applied to construct the geometric quantum gates against certain local noises \cite{Zhu2005}. Adiabatic geometric quantum computation (AGQC) based on the Berry phase has been proposed \cite{Zanardi1999, Pachos1999, Ekert2000, Duan2001} and first experimentally demonstrated in nuclear magnetic resonance (NMR) \cite{Jones2000}, aiming to realize high-fidelity and robust quantum gates. However, the long gate time due to the adiabatic and cyclic evolution conditions restricts the practical application of AGQC, especially in quantum systems with limited coherence time. Some approaches are proposed to overcome this problem, including the shortcut acceleration to the adiabatic evolution \cite{Demirplak2003, Berry2009, Muga2010, Masuda2010}, while these inevitably sacrifice some robustness and generally increase the control complexity. Recently, nonadiabatic geometric quantum computation (NGQC) has been theoretically proposed and experimentally implemented based on Abelian \cite{XiangBin2001,Zhu2002,Leibfried2003,Zhu2003,Du2006,Zhao2017, Chen2018,Xu2020,Zhao2021} and non-Abelian geometric phases \cite{Sjoeqvist2012,Xu2012,Abdumalikov2013,Feng2013,Zu2014, Xu2018,Liu2019,Yan2019,Zhu2019,Han2020} to break the limitation of the adiabatic condition. However, to strictly satisfy the cyclic evolution in NGQC, it usually requires at least $\pi$-pulse time consumption to construct a geometric gate, so there is still no advantage in operation time compared to conventional dynamical gates. Meanwhile, the increase in time consumption will also be accompanied by cumulative disturbances from systematic errors, making the robust advantage of the geometric gate displays ambiguous in experiments.

To reduce the gate-operation time and release the restriction of the cyclicity in the design of geometric gates \cite{Friedenauer2003}, some theoretical schemes based on nonadiabatic but noncyclic geometric evolution have recently been proposed \cite{Chen2020, Liu2020, Ji2021}. One of them has been experimentally implemented in a single trapped ultracold ${ }^{40} \mathrm{Ca}^{+}$ ion \cite{Zhang2021}, in which a special single-qubit geometric gate has demonstrated its error-resilient feature. But the experimental verification of short-time and error-resilient features for a set of universal geometric gates is still lacking, especially for the simultaneous suppression of different types of errors.

Here, we experimentally implement the noncyclic and nonadiabatic (NCNA) geometric quantum computation in a superconducting quantum circuit. The method we adopted to construct NCNA geometric gates is reverse engineering, which purposefully determines the Hamiltonian for the system to generate noncyclic geometric evolution paths \cite{Chen2022}. In our experiment, a set of universal and short-time single-qubit NCNA geometric gates including \cite{Nielsen2012} $\pi/8$ gate ($T$), Phase gate ($S$), and Hadamard gate ($H$) are realized, and their high fidelities are characterized via randomized benchmarking (RB). Remarkably, we also experimentally demonstrate the strong resistance of our universal single-qubit NCNA geometric gates to both the Rabi frequency error and qubit frequency shift-induced error. Finally, we implement the nontrivial two-qubit geometric operation using parametric modulation \cite{McKay2016,Ganzhorn2020, Han2020a} to generate maximally entangled Bell states.

\begin{figure}[tbp]
  \includegraphics{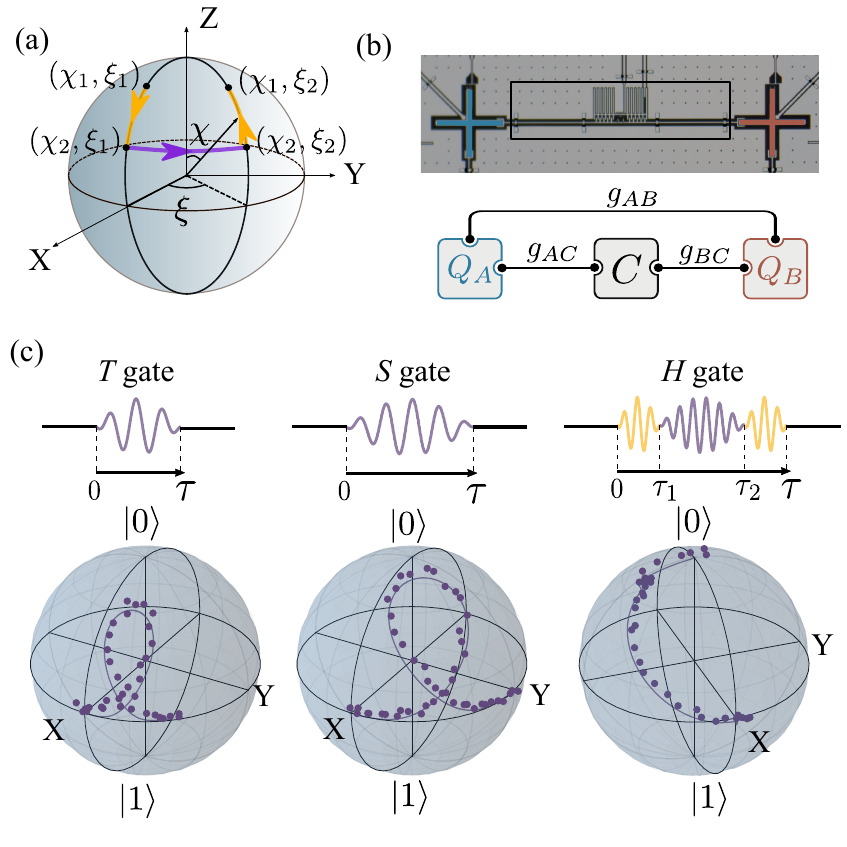}
  \caption{Single-qubit NCNA geometric gates. (a) The noncyclic evolution path of state vector $|\psi_{+}(t)\rangle$ with Bloch representation to realize NCNA geometric gates. (b) Sketch of a two-qubit system with a coupler. $ Q_A $ (blue) and $ Q_B $ (red) are directly coupled with an effective coupling strength $ g_{AB} $ and the coupling strength between the coupler $ C $ (black) and $ Q_A $ ($ Q_B $) is $ g_{AC} $ ($ g_{BC}$). (c) The experimental pulses to realize NCNA geometric $T$, $S$, and $H$ gates and the corresponding evolution trajectories with specific initial states.}
  \label{fig1}
\end{figure}

\begin{figure}[tbp]
  \includegraphics[width=8.6cm]{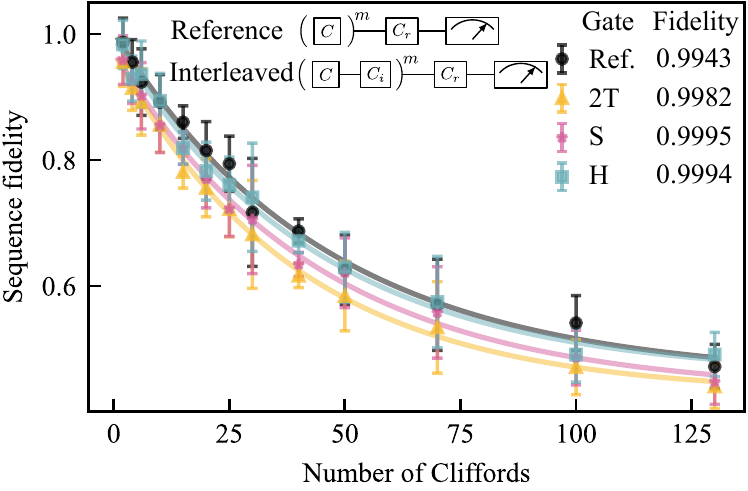}
  \caption{Clifford-based RB of single-qubit NCNA geometric gates. The inset is the sequence of gates from the Clifford group for the reference RB and interleaved RB. Sequence fidelities are functions of the number of Cliffords and the exponential decay curves give fidelities of NCNA gates: the $ 2T $, $ S $, and $ H $ gates. }
  \label{fig2}
\end{figure}

We first briefly elucidate the theoretical proposal \cite{Chen2022} of constructing NCNA geometric gates in the superconducting qubit. With $ \hbar = 1 $, a general Hamiltonian for a two-level system is
\begin{equation}
	\mathcal{H}(t)=\frac{1}{2}\begin{pmatrix}
		-\Delta(t) & \Omega(t) e^{-i \phi(t)} \\
		\Omega(t) e^{i \phi(t)} & \Delta(t)
	\end{pmatrix},
	\label{Ham1}
\end{equation}
where $ \Omega(t) $ and $ \phi(t) $ are the time-dependent amplitude and phase of the driving microwave field, respectively; $ \Delta(t) = \omega_q - \omega_m$ is the time-dependent detuning between the qubit transition frequency and the frequency of a microwave field. According to the Lewis-Riesenfeld invariant methods
\cite{Lewis1969,Chen2011,Ruschhaupt2012},
we can choose a set of orthogonal states as $|\psi_{+}(t)\rangle = e^{i f_{+}(t)}[\cos \frac{\chi(t)}{2}|0\rangle + \sin \frac{\chi(t)}{2} e^{i \xi(t)}|1\rangle]$ and $|\psi_{-}(t)\rangle = e^{i f_{-}(t)}[\sin \frac{\chi(t)}{2} e^{-i \xi(t)}|0\rangle-\cos \frac{\chi(t)}{2}|1\rangle]$ in which $ f_{+}(t)=f_{-}(t)=\gamma $ is regarded as a global phase, and $ \chi(t) $ and $ \xi(t) $ represent the polar and azimuthal angles on a Bloch sphere respectively. To realize our NCNA geometric gate, the entire noncyclic evolution path composed of three path segments needs to be utilized, as denoted in Fig.~\ref{fig1}(a). We here take the evolution details of state vector $|\psi_{+}(t)\rangle $ as an illustration: first, it evolves along the longitude line from the initial point $(\chi_1,\xi_1)$ to $(\chi_2,\xi_1)$ at time $t=\tau_1$, with null accumulation of the global phase; next the state evolves along the latitude line from $(\chi_2,\xi_1)$ to $(\chi_2,\xi_2)$ at time $t=\tau_2$; the third path is similar to the reverse of the first path, which is from $(\chi_2,\xi_2)$ to the final point $(\chi_1,\xi_2)$. Among them, after strictly eliminating the dynamical phase existing in the middle segment by setting $\int^{\tau_2}_{\tau_1}\Delta(t)dt =(\xi_1-\xi_2)\sin^2\chi_2$, the accumulated geometric phase can be obtained as
\begin{equation} \label{gammaG}
\gamma_g=-\frac{1}{2}\int^{\tau}_0 \dot{\xi}(t)	[1-\cos\chi(t)]dt=-\frac{1}{2}(\xi_2-\xi_1)(1-\cos\chi_2),
\end{equation}
which is exactly half of the solid angle enclosed by the noncyclic evolution path and its geodesic connecting the initial point $(\chi_1,\xi_1)$ and final point $(\chi_1,\xi_2)$. Based on these, the corresponding Hamiltonian parameter $\phi(t)$ and the pulse area associated with $\Omega(t)$ can then be reverse-engineered in these three segments $ t\in [0,\tau_1]$, $[\tau_1,\tau_2] $ and $ [\tau_2,\tau] $ as
\begin{eqnarray} \label{para}
&&\!\!\!\phi(t)=\xi_{1}+\frac{\pi}{2}, \quad \ \
\frac{1}{2}\int_{0}^{\tau_{1}}\Omega(t)dt = \frac{1}{2}(\chi_{2} - \chi_{1}),
\notag\\
&&\!\!\!\phi(t)=\xi(t)+\pi, \quad
\frac{1}{2}\int_{\tau_{1}}^{\tau_{2}}\Omega(t)dt=\frac{1}{4} \left(\xi_{2}-\xi_{1}\right)\sin (2\chi_{2}),
\notag\\
&&\!\!\!\phi(t)=\xi_{2}-\frac{\pi}{2}, \quad \ \
\frac{1}{2}\int_{\tau_{2}}^{\tau}\Omega(t)dt= \frac{1}{2}(\chi_{2}-\chi_{1}),
\end{eqnarray}
with detuning $ \Delta(t)=0$, $-\left(\xi_{2}-\xi_{1}\right) \sin ^{2} \chi_{2}/\left(\tau_{2}-\tau_{1}\right)$, $0$, where $ \xi(t)= \xi_{1}-\int_{\tau_{1}}^{t} \Delta\left(t^{\prime}\right) dt^{\prime}+\cot \chi_{2} \int_{\tau_{1}}^{t} \Omega\left(t^{\prime}\right) dt^{\prime}$. In this way, the resulting evolution operator is given by
\begin{eqnarray} \label{overallU}
U(\tau)=\left(
\begin{array}{cccc}
(c_{\gamma'} \!+\!is_{\gamma'} c_{\chi_1}) e^{-i \xi_-} & is_{\gamma'} s_{\chi_1} e^{-i \xi_+} \\
is_{\gamma'} s_{\chi_1} e^{i \xi_+} & (c_{\gamma'} \!-\!is_{\gamma'} c_{\chi_1}) e^{i \xi_-}
\end{array}
\!\!\right),
\end{eqnarray}
where $c_j=\cos j$, $s_j=\sin j$, $\xi_\pm=[\xi_2\pm \xi_1]/2$ and $\gamma'=\gamma_g+\xi_-$. We find that arbitrary NCNA geometric gates can be realized by setting parameters $\chi_1$, $\xi_{1,2}$, and $\gamma'$.

\begin{figure}[tbph]
  \includegraphics{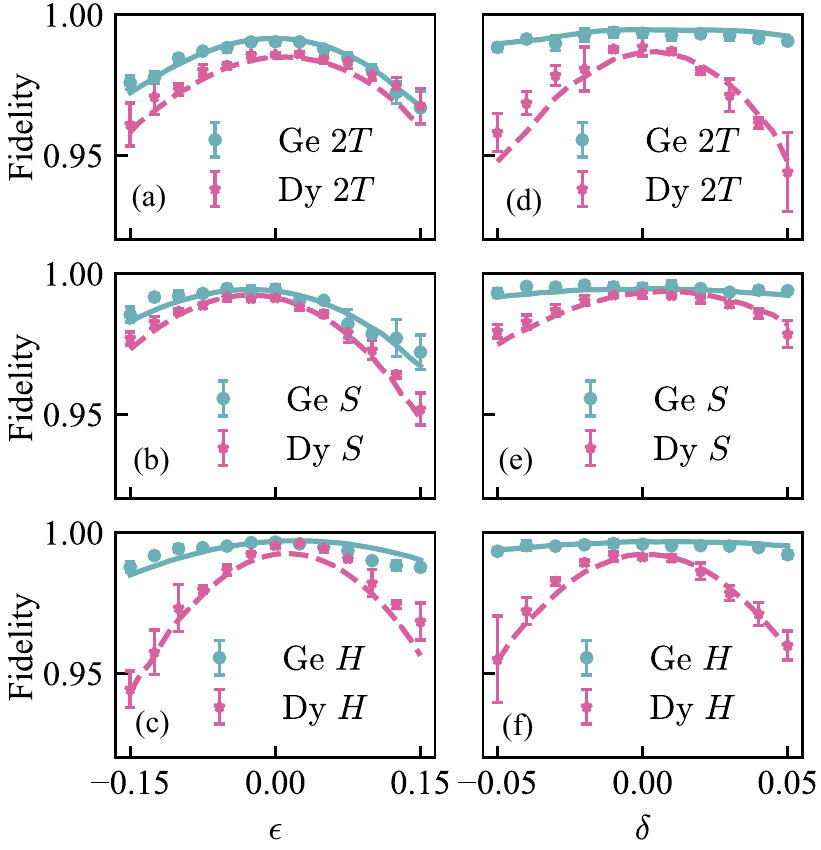}
  \caption{Error-resilient feature of NCNA geometric gates. (a)-(c) The experimental sequence fidelities as functions of the Rabi frequency error $ \epsilon $ with the interleaved RB of single-qubit geometric and dynamical gates: (a) $2T$, (b) $S$ and (c) $H$ are realized. (d)-(f) The experimental sequence fidelities as functions of the qubit frequency shift-induced error $ \delta $ with the interleaved RB of single-qubit gates: (d) $2T$, (e) $S$ and (f) $H$ are realized. The experimental results are consistent with the numerical simulation results using QuTiP \cite{Johansson2012, Johansson2013}, and the teal solid line and Fuschia dashed line represent the numerical simulation of geometric and dynamical gates, respectively.}
  \label{fig3}
\end{figure}

Our experiment of the above NCNA geometric scheme is performed in a superconducting quantum circuit containing four tunable grounded transmon qubits and four tunable floating couplers with cross-shaped capacitors \cite{Sete2021}. Only two qubits $ Q_A $ and $ Q_B $ with a coupler $ C $ are used in this experiment, with a sketch of the coupler system shown in Fig. \ref{fig1}(b). The details of the sample are introduced in the Supplementary Material \cite{Supp}. We first perform a set of universal single-qubit geometric gates, including $ \pi/8 $ gate ($T$), Phase gate ($ S $), and Hadamard gate ($ H $) on the sweet spot of $ Q_A $ to demonstrate their high fidelity and error-resilient features, in which the envelope of each pulse is a truncated Gaussian pulse with DRAG (Derivative Reduction by Adiabatic Gate) procedure \cite{Motzoi2009, Gambetta2011, Wang2018} to suppress the leakage error. To realize NCNA geometric $T$ and $S$ gates, the parameters in Eq. \ref{para} are set as $ \xi_{2}-\xi_{1} = 9\pi/4$ and $5\pi/2$ respectively, with the same $\gamma^\prime=\pi$, where we choose $\chi_{1}=\chi_{2}$ to ensure that the operation time consumed is the shortest. In addition, for the NCNA geometric $H$ gate, we set $\chi_{1}=\pi/2$, $\gamma^{\prime}=\pi/4$ and $ \xi_{2}-\xi_{1}=(2 n+1) \pi $, where $ \xi_{1} = \pi/2 $ and $ n $ is an integer. To optimize the total pulse area which corresponds $n$, we determine $n = 0$ in practice, with the limitation of sampling rate and output voltage of the arbitrary waveform generators (AWGs). Using these NCNA geometric gates obtained, we implement the geometric evolution control for the special initial states and their corresponding evolution trajectories as shown in Fig. \ref{fig1}(c). In this case, the pulse areas $\frac{1}{2}\int^{\tau}_0 \Omega(t) dt$ for these NCNA geometric $T$, $S$ and $H$ gates are about $0.46\pi$, $0.60\pi$, and $0.38\pi$, which are evidently smaller compared to the corresponding dynamical gates that are $0.625\pi $, $0.75\pi$, and $0.75\pi$.

Due to the smaller pulse area involved, the NCNA geometric scheme has an advantage in operating time compared to the conventional dynamical gate and single-loop geometric gate \cite{Xu2020}, enabling higher gate fidelity and also inheriting geometric error-resilient features. In this letter, both the advantages are characterized via the Clifford-based RB \cite{Knill2008,Magesan2011,Magesan2012}. The experimental sequences of the reference RB and interleaved RB are shown in the inset of Fig. \ref{fig2}. From the reference and interleaved RB, we obtained the depolarizing parameter $p_{\text{ref}} $ and $p_{\text{itl}}$ by fitting the experimental results shown in Fig. \ref{fig2} with $ F = Ap^m+B $, where $ A $ and $ B $ are constants that absorb preparation and measurement errors, and $ m $ is the number of Clifford gates. For the single qubit gate, the reference gate fidelity is $ F_{\text{ref}}=1-(1-p_{\text{ref}})(d-1)/d/1.875 = 0.9943$ with $ d=2 $ and the fidelities $ F_{\text{itl}} = 1-(1-p_{\text {itl}} / p_{\text {ref}})(d-1) / d $ of interleaved $ 2T $, $ S $ and $ H $ gates are $ 0.9982 $, $ 0.9995 $, and $ 0.9994 $, respectively (since the $ T $ gate is not a Clifford generator, we apply two $ T $ gates in series to demonstrate the fidelity of the NCNA geometric $T$ gate \cite{Barends2014}).

Furthermore, to verify the gate robustness of the NCNA geometric scheme, we next consider the error-affected Hamiltonian as follows:
\begin{gather}
	\mathcal{H}(t)=\frac{1}{2}\left(\begin{array}{cc}
		-\left(\Delta+\delta \Omega_{\mathrm{m}}\right) & (1+\epsilon) \Omega(t) e^{-i \phi(t)} \\
		(1+\epsilon) \Omega(t) e^{i \phi(t)} & \left(\Delta+\delta \Omega_{\mathrm{m}}\right)
	\end{array}\right),
\end{gather}
where $\epsilon$ and $\delta$ represent the pulse amplitude (Rabi frequency) error and qubit frequency shift-induced error respectively, and $\Omega_m$ is the maximum of $\Omega(t)$. In the experiment, these errors are generated by the designed microwave pulses. We continue to compare the robustness of the error-affected NCNA geometric $2T$, $S$, and $H$ gates with the corresponding dynamical counterparts by using interleaved RB. To numerically simulate the RB results, we compute all propagators of the system Hamiltonian corresponding to single-qubit gates considering the relaxation and dephasing time, and we also consider the second excited state to calculate the population leakage. Similar to the experimental RB procedure, the propagators are applied to the state density matrix in the vector representation. Then the fidelities of numerical simulation can be finally obtained. As shown in Fig. \ref{fig3}, our experimental and numerical results demonstrate the suppression effects of NCNA geometric $2T$, $S$, and $H$ gates on both the Rabi frequency error and qubit frequency shifted-induced error. Thus, we experimentally implement a set of universal single-qubit NCNA geometric gates, which outperform dynamical gates comprehensively in the gate robustness and gate time.

\begin{figure}[tbp]
	\includegraphics{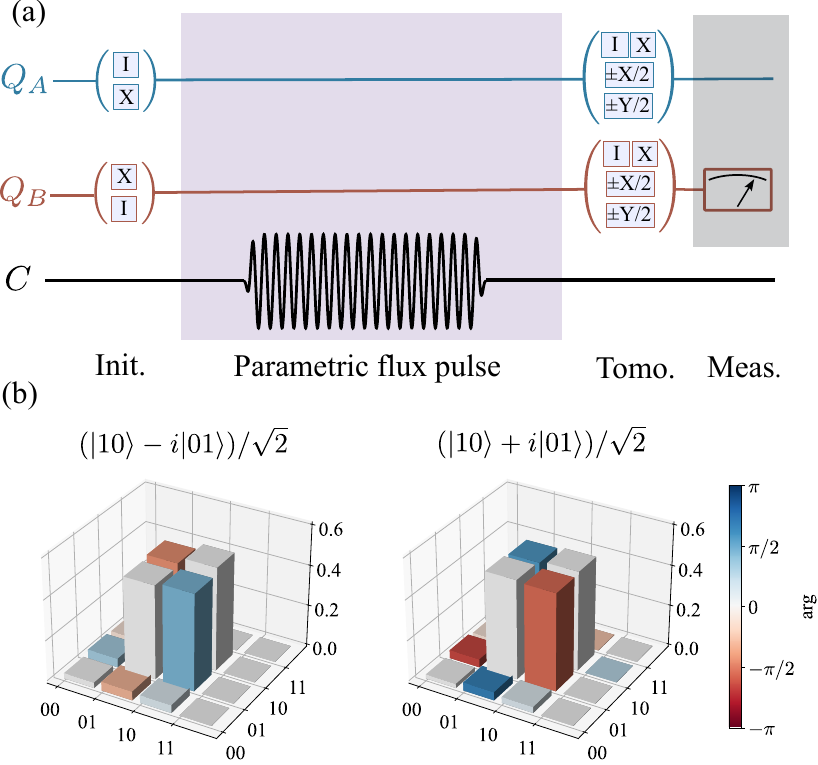}
 \caption{Bell states generated by the two-qubit NCNA geometric operation. (a) The whole pulse sequence is used to demonstrate Bell states. The system can be initialized in $ |01\rangle $ or $ |10\rangle $ and then we perform a well-designed parametric flux pulse. Finally, we use single-qubit gates $ I $, $ X $, $ \pm X/2 $ and $ \pm Y/2 $ to implement a joint dispersive readout using the resonator of $ Q_B $. (b) The density matrix of Bell states $ (|10\rangle-i|01\rangle) / \sqrt{2} $ and $ (|10\rangle+i|01\rangle) / \sqrt{2} $ generated by the two-qubit NCNA geometric operation. The heights and colors of bars in the histogram represent the corresponding amplitudes and arguments of density matrix elements.}
	\label{fig4}
\end{figure}

The implementation of quantum computation also involves the entangled interactions between qubits.
In this letter, we experimentally demonstrate that the NCNA geometric scheme can also be applied to two-qubit manipulation, in which the construction of two-qubit entangled states by using the parametric modulation \cite{McKay2016, Ganzhorn2020, Han2020a} is taken as an example. The coupling model in our experiment consisting of two qubits $Q_{A,B}$ and a coupler $C$ is shown in Fig. \ref{fig1} (b). $Q_A$ and $Q_B$ are biased at the operating spot and the sweet spot, respectively, and the coupler $C$ is modulated using a parametric pulse in the form of $\phi(t)=\phi_{\mathrm{dc}}+\epsilon_p\cos(\omega_p t+\phi_p)$, in which $ \phi_{\mathrm{dc}} $, $ \epsilon_p $, $ \omega_p $, and $ \phi_p $ are the DC flux bias, modulation amplitude, frequency, and phase, respectively. After neglecting the high-order oscillating terms and applying the unitary transformation, the final effective Hamiltonian within the subspace $ \{|01\rangle, |10\rangle\} $ can be written as
\begin{equation}
   H_{\mathrm{eff}}=\frac{1}{2}\left(\begin{array}{cc}
     -\Delta^{\prime} & g_{\mathrm{eff}} e^{-i(\eta t+\varphi)} \\
     g_{\mathrm{eff}} e^{i(\eta t+\varphi)} & \Delta^{\prime}
   \end{array}\right),
 \label{Heff}
\end{equation}
where $ g_{\mathrm{eff}} $ and $\eta t+\varphi$ are the effective coupling strength and the time-dependent phase generated by modulation pulses, a detailed derivation can be found in the supplementary material \cite{Supp}, while the detuning $ \Delta^{\prime} $ is defined as $\omega_p - (\omega_{01} -\omega_{10})$. 

The above Hamiltonian form is the same as the single-qubit one $\mathcal{H}(t)$, thus we can determine Hamiltonian parameters of Eq. \ref{Heff} according to Eq. \ref{para}, to implement an arbitrary geometric operation in the subspace spanned by $ \{|01\rangle, |10\rangle\} $, while the experimental pulse sequence is shown in Fig. \ref{fig4}(a). As a result, extending to the whole two-qubit subspace, the entangled two-qubit geometric manipulation is realized. Next, we proceed to implement high-fidelity entangled Bell states preparation based on this two-qubit geometric manipulation, in which we set $\chi_1=\chi_{2}=0.789 $ and $ \xi_{2}-\xi_{1}= 3\pi/2 $. The modulation phase $ \phi_p = \pi $ and the amplitude $ \epsilon_p $ are designed to induce the effective strength $ g_{\mathrm{eff}} = 3.96 $~MHz. The frequency of modulation pulse become $ \omega_{p}= \omega_{01} -\omega_{10} + g_{\mathrm{eff}} \cot{\chi_2} = 163.7 $~MHz. To generate high-fidelity Bell states, the duration of the longitudinal waveform is calibrated through the population in the subspace $ \{|01\rangle, |10\rangle\} $ \cite{Supp}. We carefully design the pulse area with the width of the rising, square, and falling pulses are $ 10 $~ns, $ 78 $~ns, and $ 10 $~ns respectively. Here, we perform tow-qubit state tomography using a joint dispersive readout \cite{Filipp2009, Chow2010} to detect Bell states shown in Fig. \ref{fig4}(b). The two-qubit system is initialized in the state $ |10\rangle $ ($ |01\rangle $) and then the pulse of the entangled operation is applied to the flux of the coupler $C$. The Bell states can be restructured by implementing an overcomplete raw measurement involving different combinations of single-qubit gates $I$, $X$, $\pm X/2$, and $\pm Y/2$ on $Q_A$ ($Q_B$), respectively \cite{Supp}. The generated Bell states are $ (|10\rangle-i|01\rangle) / \sqrt{2} $ and $ (|10\rangle+i|01\rangle) / \sqrt{2} $ with fidelities $ 97.89\% $ and $ 98.07\% $ when the corresponding initial states are $ |10\rangle $ and $ |01\rangle $. The comparisons of experimental and ideal Bell states are shown in Ref. \cite{Supp}.

In summary, we experimentally implement the short-time and high-fidelity NCNA geometric gates in the superconducting circuit, and also experimentally demonstrate that these geometric gates are more robust to both the Rabi frequency error and qubit frequency shift-induced error compared to the conventional dynamical gates. In addition, the approach can be generalized to two-qubit manipulation to generate the Bell states. Therefore, the NCNA geometric gates are a promising candidate for fast, high-fidelity, and robust universal quantum operations.

\begin{acknowledgments}
This work was partly supported by the Key R\&D Program of Guangdong Province (Grant No. 2018B030326001), NSFC (Grant No. 12074179, No. 11890704, and No. U21A20436), NSF of Jiangsu Province (Grant No. BE2021015-1), and Innovation Program for Quantum Science and Technology (2021ZD0301700).
\end{acknowledgments}

\section{Reference}

\providecommand{\noopsort}[1]{}\providecommand{\singleletter}[1]{#1}%

\end{document}